\documentclass[prl,aps,amssymb,showpacs,twocolumn]{revtex4}
\usepackage{amsmath}
\usepackage{amssymb}
\usepackage{amsthm}
\usepackage{amsfonts}
\usepackage{enumerate}
\usepackage{latexsym}
\usepackage{graphicx}

\newcommand{\beq}{\begin{equation}}
\newcommand{\eneq}{\end{equation}}

\input{epsf}

\begin{document}

\tolerance 10000

\newcommand{\vk}{{\bf k}}


\title{Fractional Quantum Hall States and Jack Polynomials}

\author{B. Andrei Bernevig$^{1,2}$ and F.D.M. Haldane$^2$}

\affiliation{ } \affiliation{$^1$Princeton Center for Theoretical
Physics, Princeton, NJ 08544} \affiliation{$^2$ Department of
Physics, Princeton University, Princeton, NJ 08544}

\begin{abstract}
We describe an occupation-number-like picture
of Fractional Quantum Hall (FQH) states in
terms of  polynomial wavefunctions  characterized by a dominant
occupation-number configuration.
The bosonic variants of single-component abelian and non-abelian
FQH states are modeled by Jacks (Jack  symmetric polynomials), characterized
by dominant  occupation-number configurations
satisfying a generalized Pauli principle. In
a series of well-known Quantum Hall states,
including the Laughlin, Read-Moore, and
Read-Rezayi, the Jack polynomials naturally implement
a ``squeezing rule'' that constrains allowed
configurations to be restricted to those obtained by
squeezing the dominant configuration. The Jacks describing uniform FQH
states satisfy a highest-weight
condition, and  a clustering
condition which can be generalized to describe quasiparticle states.
\end{abstract}

\date{July 23, 2007}


\maketitle

The Laughlin wavefunction\cite{laughlin1983} has provided the key
to understanding the physics of the fractional quantum Hall (FQH) effect:
it accurately models the simplest abelian FQH states
and is the building block of
model wavefunctions for more general states,
both abelian, and (using cluster projections) non-abelian ones, such as the
Moore-Read\cite{moore1991} and Read-Rezayi\cite{read1999} states.
Apart from trivial (gaussian) factors which we will drop,
such model wavefunctions are conformally-invariant multivariable polynomials
$\psi(z_1,\ldots,z_N)$;
despite their explicit availability,
analytic calculations of correlation functions and other physical
properties have not so far been possible because of the intractability of
their expansions in the non-interacting basis
of occupation number states (Slater determinants or monomials).
As a result, quantitative study of these states
has relied on exact diagonalization and other numerical methods
\cite{haldane1983}.

The simplest physically-relevant model FQH states are antisymmetric
polynomials, describing spin-polarized electrons in a partially-filled
Landau level with no internal ``pseudospin'' degrees of freedom, but
it is also useful to study symmetric (bosonic) FQH wavefunctions
from which they are obtained by multiplication by odd powers of the
Vandermonde determinant.

In this Letter, we describe a unified occupation-basis framework for
the description of many model one-component  FQH states in terms of
the Jack symmetric polynomial(s) (``Jack(s)'')\cite{stanley1989}.
The Jacks naturally implement a type of ``generalized Pauli
principle'' on a generalization of Fock spaces for abelian and
non-abelian fractional statistics\cite{haldaneUCSB2006}. We note
that (bosonic) Laughlin, Moore-Read, and Read-Rezayi wavefunctions
(as well as others, such as the state Simon \textit{et
al.}\cite{simon2006} have called the ``Gaffnian'') can be explicitly
written as Jack symmetric polynomials, which have known (recursively
defined) expansions in monomials (free boson occupation number
states), and have rich algebraic properties. These uniform FQH
condensate wavefunctions can be obtained by requiring that a Jack
simultaneously obeys highest-weight (HW, absence of quasiholes) and
lowest-weight (LW, absence of quasiparticles) conditions.

\begin{figure}
\includegraphics[width=3.2in, height=1.9in]{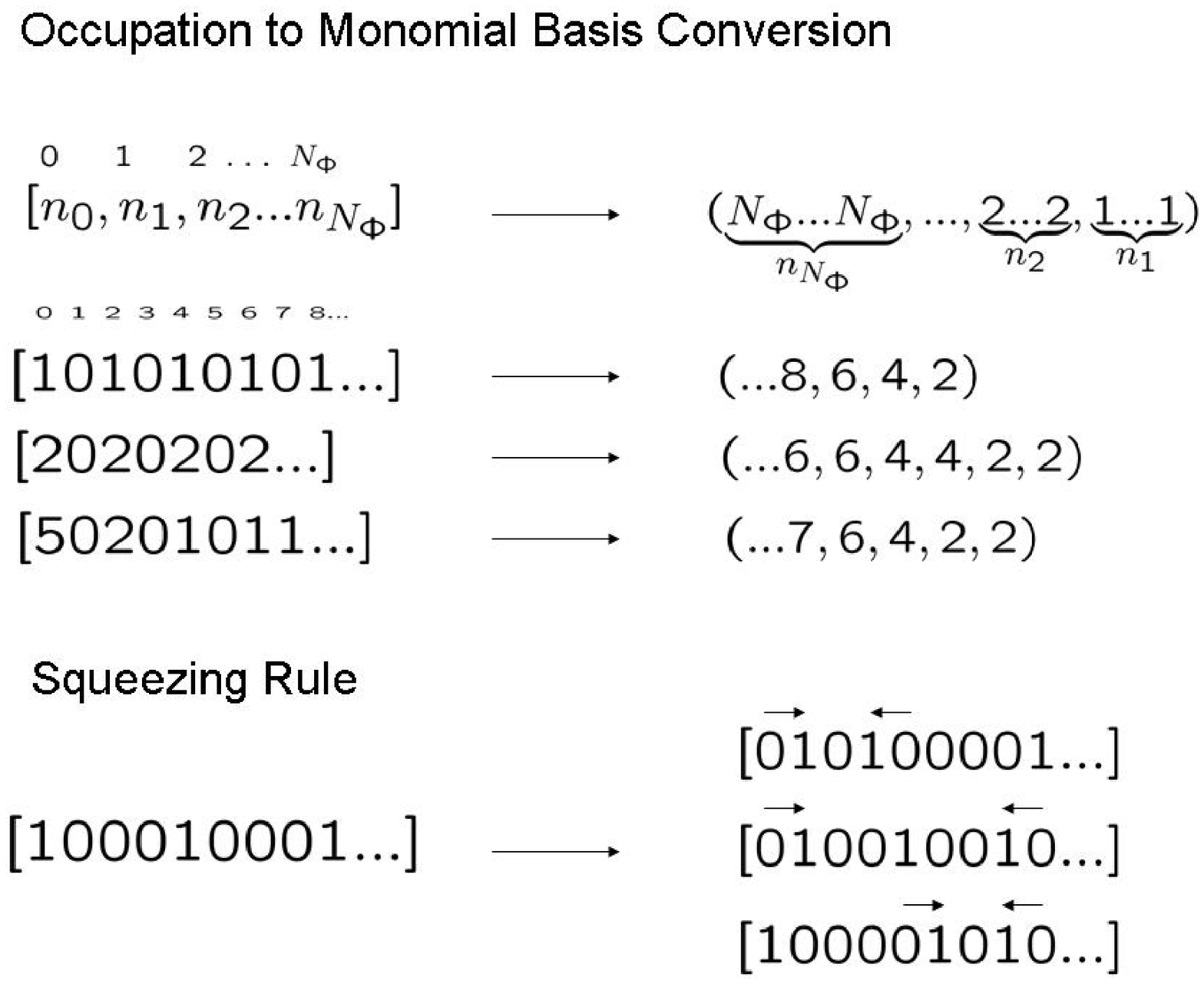}
\caption{Occupation to monomial basis conversion and squeezing rule
example}\label{occupation}
\end{figure}

Jacks $J^{\alpha}_{\lambda}(z)$ are symmetric polynomials in $z$
$\equiv$ $\{z_1,z_2,\ldots ,z_N\}$, labeled by a partition $\lambda$
with length $\ell_{\lambda} \le N$, and a parameter $\alpha$;
$\lambda$ can be represented as a (bosonic) occupation-number
configuration $n(\lambda)$ = $\{n_m(\lambda),m=0,1,2,\ldots\}$ of
each of the lowest Landau level (LLL) orbitals with angular momentum
$L_z = m \hbar$ (see Fig[\ref{occupation}]), where, for $m > 0$,
$n_m(\lambda)$ is the multiplicity of $m$ in $\lambda$. When
$\alpha$ $\rightarrow$ $\infty$, $J^{\alpha}_{\lambda}$
$\rightarrow$ $m_{\lambda}$, which is the monomial wavefunction of
the free boson state with occupation-number configuration
$n(\lambda)$; a key property of the Jack $J^{\alpha}_{\lambda}$  is
that its expansion in terms of monomials only contains terms
$m_{\mu}$ where $\mu$ $\le$ $\lambda$, where $\mu$ $<$ $\lambda$
means the partition $\mu$ is \textit{dominated} by
$\lambda$\cite{stanley1989}. Jacks are also eigenstates of a
Laplace-Beltrami operator $\mathcal H_{\rm LB}(\alpha)$ given by
\begin{equation}
\sum_i \left ( z_i \frac{\partial}{\partial z_i} \right )^2
+ \frac{1}{\alpha}\sum_{i<j}\frac{z_i+z_j}{z_i-z_j}
\left (
z_i \frac{\partial}{\partial z_i} -
z_j \frac{\partial}{\partial z_j} \right ).
\end{equation}

We note that the bosonic Laughlin state $\psi^{(r)}_{\rm L}$
at filling $\nu=1/r$, $r$ even, is a Jack polynomial:
\begin{equation}
\psi^{(r)}_{\rm L} =\prod_{i<j}^N (z_i-z_j)^r =
J^{\alpha_{1,r}}_{\lambda^0(1,r)}(z), \quad
 \alpha_{k,r}\equiv -{\textstyle \frac{k+1}{r-1}},
\label{jacklaughlin}
\end{equation}
which is the $k$ = 1 case of a Jack defined for any positive integer
$k$ so that $N$ = $k\bar N$, and $n_m(\lambda^0(k,r))$ = $k$ for $m$
= $(j-1)r$, $j$ = $1,2,\ldots,\bar N$, with $n_m$ = 0 otherwise.
Note that $\lambda^0(k,r)$ is the ``($k$,$r$,$N$)-admissible''
partition \cite{feigin2002} that minimizes $|\lambda|$ $\equiv$ $M$
$\equiv$ $\sum_m mn_m$ at fixed $N$ ($\lambda$ is
``($k$,$r$,$N$)-admissible'' if $n(\lambda)$ obeys a ``generalized
Pauli principle'' where, for all $m\ge 0$, $\sum_{j=1}^rn_{m+j-1}$
$\le$ $k$, so $r$ consecutive ``orbitals'' contain no more than $k$
particles). Note that here the Jack parameter $\alpha_{k,r}$ is a
\textit{negative rational}; study of symmetric Jacks of this type
was recently initiated in Ref.\cite{feigin2002}. Earlier work
generally assumes $\alpha$ is a positive real (Jacks with real
$\alpha >0$ and unrestricted $\lambda$ occur in the solution of the
integrable Calogero-Sutherland model\cite{sutherland1971}).
Non-symmetric Jack-polynomials can also describe spin-FQHE
wavefunctions such as Halperin and Haldane-Rezayi
\cite{kasatani2006}.

It is straightforward to see that $\psi^{(r)}_{\rm L}$ is a Jack:
it has the obvious property that it is
annihilated by  operators
\begin{equation}
D_i^{L,r} =\frac{\partial}{\partial z_i} - r \sum_{j(\ne i)}'
\frac{1}{z_i -z_j};\;\;\;\; D_i^{L,r} \psi^{(r)}_{\rm L}=0.
\end{equation}
\noindent It is then also annihilated by the combination
$\sum_i z_i D_i^{L,1} z_i D_i^{L,r}$, which is equal to
$\mathcal H_{\rm LB}(\alpha_{1,r})$ minus a constant (found by
direct calculation to be $\frac{1}{12}rN(N-1)(N+1 + 3r(N-1)$), so
$\psi^{(r)}$ is an eigenstate of $\mathcal H_{\rm
LB}(\alpha_{1,r})$. It is now easy to identify the dominant
configuration $n(\lambda^0_{1,r})$, and verify that the eigenstate
is non-degenerate, confirming Eq.(\ref{jacklaughlin}). For $r$ = 2,
this also follows implicitly from Ref.\cite{feigin2002} where
it was shown that the set of Jacks with parameter $\alpha_{k,2}$ and
($k$,$2$,$N$)-admissible $\lambda$ are a basis for the space of
symmetric polynomials that vanish when $k+1$ variables $z_i$
coincide. The space of symmetric polynomials space can be divided
into subspaces of fixed $M$ = $|\lambda|$, and for $|\lambda|$ =
$\lambda^0_{k,r}$, there is a \textit{single}
($k$,$r$,$N$)-admissible partition, so a polynomial with the
appropriate properties is unique, and must be a Jack.

It is useful to identify the ``dominance rule'' (a partial ordering
of partitions $\lambda > \mu$) with the ``squeezing
rule''\cite{sutherland1971} that connects configurations
$n(\lambda)$ $\rightarrow$ $n(\mu)$: ``squeezing'' is a two-particle
operation that moves a particle from orbital $m_1$ to $m_1'$ and
another from $m_2$ to $m_2'$, where $m_1 < m_1' \le m_2' < m_2$, and
$m_1+m_2$ = $m_1'+m_2'$; $\lambda > \mu$ if $n(\mu)$ can be derived
from $n(\lambda)$ by a sequence of ``squeezings'' (see
Fig.\ref{occupation}). This means that when model FQH wavefunctions
equivalent to Jacks are expanded in basis of occupation number
states, only configurations obtained by ``squeezing''' from a
dominant configuration will be present (this crucial property
persists in fermionic model FQH wavefunctions given by
the product of a Jack with a power of the Vandermonde determinant).

Jacks can be normalized so that
\begin{equation}
J^\alpha_{\lambda} = m_\lambda + \sum_{\mu<\lambda} v_{\lambda
\mu}(\alpha) m_\mu .
\end{equation}
The coefficients $v_{\lambda \mu}(\alpha)$ are (recursively) known
\cite{lapointe2000}; they are finite and real positive for real
$\alpha > 0$, and are holomorphic functions of $\alpha$ except for
poles at a ($\lambda$,$\mu$)-dependent set of negative rational
values\cite{feigin2002}. Feigin \textit{et.al.}\cite{feigin2002}
proved that for  the ($k$,$r$,$N$)-admissible partitions,
$v_{\lambda \mu}(\alpha)$ is analytic at $\alpha_{k,r}$, and the set
of admissible Jacks with this parameter form a basis of a
differential ideal $I_N^{k,r}$ in the space of symmetric
polynomials. This requires that $(k+1)$ and $(r-1)$ (but \textit{not
necessarily} $k$ and $r$) be coprime. For the case $r=2$, and $k$
integer, these polynomials are a basis for the $\nu$ = $k/r$ = $k/2$
bosonic non-abelian Read-Rezayi FQH states with quasiholes, with
special cases $k$ = 1 (Laughlin state) and $k$ = 2 (Moore-Read
state). By multiplying these wavefunctions by $\psi^{(m)}_{\rm L}$,
this generalizes to the $\nu$ = $k/(km+2)$ Read-Rezayi states, and
reproduces the ``generalized Pauli principle'' exclusion statistics
structure found empirically in numerical studies by one of
us\cite{haldaneUCSB2006}.

The $\nu$ = $1/r$ Laughlin state is a Jack-polynomial with parameter
$\alpha_{1,r}$ and $n(\lambda)$ = $[10^{r-1}10^{r-1}\ldots ]$, where
``$0^{r-1}$'' means a sequence of $r-1$ ``empty orbitals''. A basis
of one-quasihole states can similarly be shown to be given by the
Jack with $n(\lambda)$ = $[10^{r-1}1\ldots 0^{r-1}10^r10^{r-1}
\ldots]$, where there is a single extra ``empty orbital''. These
states all have different $M$, and hence are orthogonal, and form a
multiplet. A linearly-independent basis of two-quasihole state is
given by Jacks with the same $\alpha$ and two extra ``empty
orbitals''' in $n(\lambda)$.   For example, at $r=2$,  two such
configurations $n(\lambda)$ and $n(\lambda')$ (with the same $M$)
are $[10100010101\ldots]$ and $[10010100101\ldots ]$.  While
$m_{\lambda}$ and $m_{\lambda'}$ are
orthogonal free-boson wavefunctions, the Jack  FQH wavefunctions
$J^{\alpha_{1,2}}_{\lambda}$ and $J^{\alpha_{1,2}}_{\lambda'}$ are
not orthogonal with respect to the usual quantum-mechanical
scalar product.
This highlights an important difference between the basis of
``admissible Jacks'' (with a ``generalized Pauli principle'') and
the ordinary free-particle basis: the pure Jack wavefunctions are
not eigenstates of a Hermitian Hamiltonian ($\mathcal H_{\rm
LB}(\alpha) $ is not Hermitian for finite $\alpha$), and are
linearly-independent but not orthogonal.
(In contrast,
Jacks with $\alpha$ real positive (and unrestricted $\lambda$) are orthogonal
with respect to a combinatorically-motivated scalar product\cite{stanley1989}
and also as  Calogero-Sutherland model wavefunctions.)

\begin{figure}
\includegraphics[width=3.2in, height=2.1in]{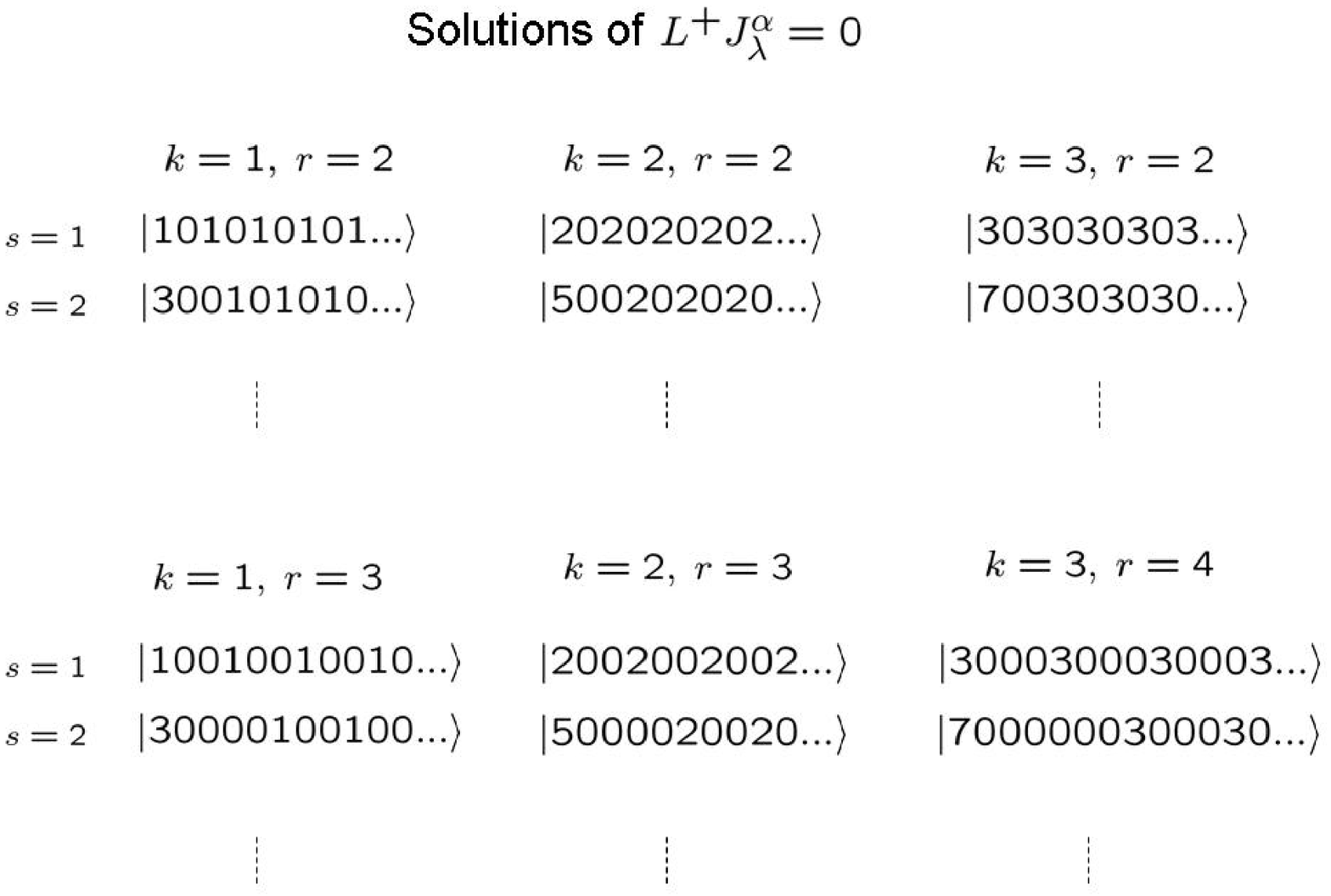}
\caption{Solutions to $L^+ J^\alpha_\lambda=$ are parametrized by
one integer, $s>0$. Only $s=1$ states are highest weight states on
the sphere, and satisfy the clustering property that they vanish as
the $r$'th power of the distance between $k+1$-particles. Other
states, related to quasiparticles, satisfy generalized clustering
conditions to be described in a future paper \label{fig2}
\cite{uslater}}\label{E0}
\end{figure}

Partitions $\lambda$ can be classified by $\lambda_1$, their largest
part.  When $J^{\alpha}_{\lambda}$ is expanded in occupation-number
states (monomials), no orbital with $m >\lambda_1$ is occupied, and
Jacks with $\lambda_1 \le N_{\Phi}$ form a basis of FQH states on a
sphere surrounding a monopole with charge
$N_{\Phi}$\cite{haldane1983}.   Uniform states on the sphere satisfy
conditions $L^+\psi$ = 0  (highest weight, HW) and $L^-\psi$ = 0
(lowest weight, LW) where $L^+$ = $E_0$, and $L^-$ =
$N_{\Phi}Z-E_2$, where $Z$ $\equiv$ $\sum_i z_i$, and $E_n$ =
$\sum_iz_i^n\partial/\partial z_i$. When both conditions are
satisfied, $E_1\psi$ $\equiv$ $M\psi$  = $ \frac{1}{2}
NN_{\Phi}\psi$. It is very instructive to find the conditions for a
Jack to satisfy the HW condition, $E_0J^{\alpha}_{\lambda}$ = 0. The
action of $ E_0$ on a Jack can be obtained from a formula due to
Lassalle\cite{lassalle1998}:  using the property that for real
$\alpha > 0$, all the $v_{\lambda}(\alpha)$ are real
positive\cite{stanley1989}, the HW condition can only be satisfied
for real $\alpha < 0$. Another condition we find is that $n_0$
$\equiv$ $N-\ell_{\lambda}$ $ > 0$ (non-zero occupancy of the $m=0$
``orbital'').  We then find a necessary (but not sufficient)
condition is
\begin{equation}
N-\ell_{\lambda} + 1 + \alpha(\lambda_{\ell}-1) = 0,
\end{equation}
where $\lambda_{\ell}$ is the smallest (non-zero) part in $\lambda$.
This imposes the following two conditions: (\textit{i}) $\alpha$ is
a negative rational, which we can choose to write as $-(k+1)/(r-1)$,
with $(k+1)$ and $(r-1)$ both positive, and relatively prime;
(\textit{ii}) $\lambda_{\ell}$ = $(r-1)s + 1$, and $n_0$ =
$(k+1)s-1$, where $s > 0$ is a positive integer.  The remaining
HW conditions require that all parts in $\lambda$ have multiplicity $k$,
so that $n(\lambda)$ =  $[n_0 0^{s(r-1)}k 0^{r-1}k 0^{r-1}k....]$,
(\textit{i.e}, the ($k$,$r$,$N$)-admissibility condition
is satisfied as an equality for orbitals $m$ $\ge$
$\lambda_{\ell}$). The case $s$ = 1 gives the FQH ground states which
also obey the LW condition, with filling factor $\nu$ = $k/r$, while
the cases $s> 1$ are intimately related to what we interpret as the
\textit{quasiparticle} (not quasihole) excitations of these $\nu$ =
$k/r$ FQH states, where $r$ = 2 corresponds to the bosonic
Laughlin/Moore-Read/Read-Rezayi sequence (see Fig.\ref{fig2}).

We will describe the quasiparticle construction elsewhere\cite{uslater},
but note that when it is
applied to the case $k=1$, $r=2$ ($\nu$ = 1/2 bosonic Laughlin
state), it reproduces the model quasiparticle state given by Jain's
projective construction\cite{jain1989}. The above derivation very
simply reproduces the admissibility conditions found in
Ref.\cite{feigin2002}, and shows that the ($k$,$r$,$N$)-admissible
Jacks that minimize $M$ at fixed $N$=$k\bar N$ are the only pure
Jacks that are acceptable FQH wavefunctions.   We also remark that
the construction of the $s>1$ sequence may prove useful\cite{uslater}
in connection with an open mathematical problem (the Cayley-Sylvester problem
of classifying polynomials with coincident roots).

We now turn our attention to the Moore-Read state\cite{moore1991}. It was
introduced
as a model for the observed $\nu =5/2$ spin-polarized FQH
($\nu$ = 1/2 in the
second LL) and is
the $m=1$ case of the $\nu=2/(2m+2))$ state
\begin{equation}
\Psi^m_{\rm MR} = \prod_{i<j} (z_i -z_j)^{m+1}
\mathop{\rm Pf}\left(\frac{1}{z_i - z_j} \right),
\end{equation}
where $\Psi^0_{\rm MR}$ is a $\nu=1$ FQH state of bosons at
$\nu=k/r$ where $k$ = $r$ = 2:
The
exclusion statistics picture of this bosonic
state is $n(\lambda^0(2,2))=[20202...]$
\cite{haldaneUCSB2006}, or (the highest density state with) not more
than $2$ particles in $2$ consecutive orbitals. It was initially
defined as the correlation function of an Ising Majorana field
$\psi_{(2,1)}=\psi(z)$ with scaling dimension $h_{2,1}=\frac{1}{2}$
in the minimal model $M(4,3)$ with $c=\frac{1}{2}$ (small indices
label degenerate fields in CFT). The correlation functions of a field
$\psi_{(m,n)}$ satisfy an  $nm$'th-order  differential equation.
This allows us to define a set of $N$ annihilation operators. For the
Pfaffian $\langle \psi(z_1) \psi(z_2)...\psi(z_N)\rangle =
\mathop{\rm Pf}\left(\frac{1}{z_i -z_j} \right)$ the annihilation operators are
\cite{belavin1984}:
\begin{equation}
D_i^{\rm Pf} =  \frac{\partial^2}{\partial z_i^2} - \sum_{j \ne i}
\frac{A_{2,1}}{z_i - z_j} \frac{\partial}{\partial z_j} - \sum_{j
\ne i} \frac{B_{2,1}}{(z_i - z_j)^2}; \label{pfdiffeq}
\end{equation}
where $A_{2,1}= 2  (2h_{2,1} +1)/3$ and $B_{2,1}=h_{2,1} A_{2,1}$.
The Pfaffian satisfies $D_i^{\rm Pf} \mathop{\rm
Pf}\left(\frac{1}{z_i-z_j}\right)=0$. According to the general
prescription for obtaining Quantum Hall wavefunctions out of CFT
correlators, the first bosonic Moore-Read state $\psi^0_{\rm MR}$ is
obtained by multiplying the Pfaffian by a Vandermonde factor:
$\psi^{(1)}_{\rm L}$. It is straightforward to transform $D_i^{\rm
Pf}$ to obtain related operators $D_i^{\rm MR}$ that annihilate
$\psi_{\rm MR}$, and show that $\sum_i D_i^{\rm MR}$ is $\mathcal
H_{\rm LB}(-3)$ plus a constant (found by direct computation to be
$- N(16 -18 N +5 N^2)/18$), which confirms that $\psi^0_{\rm MR}$ $=
$ $J^{-3}_{\lambda^0(2,2)}$.

The $\nu$ = $k/2$ bosonic
Read-Rezayi (RR) states\cite{read1999} are ``$Z_k$ parafermion states''.
The first RR state is related
to the $Z_3$ Potts model\cite{dotsenko1984} and is
annihilated by a third order differential operator. The dominant
configuration this state is
$n(\lambda^0(3,2))=[3030303...]$, or (the highest density state
with) not more than $3$ particles in $2$ consecutive orbitals. The
$Z_3$ parafermion quantum Hall state is a single Jack and
diagonalizes the second-order Laplace-Beltrami operator. The RR
$Z_k$ sequence is $\psi^0_{RR}(z)$ =
$J^{\alpha(k,2)}_{\lambda^0(k,2) }(z)$.

A bosonic state at $\nu$ = 2/3 (or a fermionic one with $\nu$ = 2/5) has
been referred to as a ``Gaffnian'' \cite{simon2006}. The dominant
configuration of the
bosonic  state  $\psi^0_{\rm G}$ is $n(\lambda^0(2,3))=[2002002002...]$, or the
highest density $(2,3)$ state. We find this state is annihilated by
(\ref{pfdiffeq}) with $h_{2,1} =3/4$, as expected, as the
wavefunction of this state is also the correlation function of  a
minimal CFT $M(5,3)$ field $\psi_{(2,1)}$ with this scaling
dimension\cite{simon2006}, and we identify $\psi^0_{\rm G}(z)$ as the $(2,3)$
vacuum  Jack $J^{\alpha_{2,3}}_{\lambda^0(2,3)} (z)$.

Instead of using the differential equations that they satisfy, it is easier to
identify the FQH states with Jacks from their clustering
properties. Information on how the Jacks vanish as $k+1$ coordinates
coincide is needed:
we verified that, for \textit{any} ($k$,$r$,$N$)-admissible $\lambda$,
if $z_1$ = $z_2$ = $\ldots $ = $z_k$ = $Z$, that
$J^{\alpha_{k,r}}_{\lambda}(z)$
has a factor $\prod_{i=k+1}^N(Z-z_i)^r$, showing how it vanishes as a cluster
of $k+1$ coincident coordinates is formed.
This agrees with the known properties of the bosonic
Laughlin/Moore-Read/Read-Rezayi states ($k \ge 1$, $r$ = 2), as well as
the ``Gaffnian'' ($k$ = 2, $r$ = 3).
For the case of the FQH ground states, where  $\lambda$
= $\lambda^0_{k,r}$, the Jacks satisfy a stronger clustering property
that relates $N$- and
$(N+k)$-particle states: for $\{z\}$ $\equiv$
$\{z_1,\ldots,z_N\}$,
\begin{equation}
\prod_{i=1}^N (Z-z_i)^r J^{\alpha_{k,r}}_{\lambda^0(k,r)}(\{z\}) =
J^{\alpha_{k,r}}_{\lambda^0(k,r)}(\{z\},Z,\ldots,Z) ,
\end{equation}
where on the RHS, $z_i = Z$ for $i$ = $N+1,\ldots, N+k$. As a
corollary, when a ($k$,$r$) Jack FQH ground state is fully
$k$-clustered, \textit{i.e.},  $z_{ki-j}$ $\rightarrow$ $Z_i$ for
$i$ = $1,\ldots ,\bar N$ and $j$ = $1,\ldots, k$, it becomes a
Laughlin state in the cluster coordinates:
\begin{equation}
J^{\alpha_{k,r}}_{\lambda^0(k,r)} (z) \rightarrow \psi^{(kr)}_L(Z_i)
.
\end{equation}
These properties show that, for these model FQH states, removing a
cluster of $k$ particles at a point $Z$ is exactly equivalent to
inserting $r$ flux quanta (or vortices) at that point (as is
well-known in the $k$ = 1 Laughlin case, and implicit in the
Read-Rezayi construction of the parafermion states as
symmetrizations over distinct pairs of Laughlin-like clusters of
particles \cite{read1999} ).

 As is obvious from their clustering property, the
($k$,$r$,$N$)-admissible Jacks also have the property that a
($k+1$)-cluster of particles cannot have relative angular momentum
less than $r$, and hence are simultaneous null states of Hermitian
operators $\hat H^{(k+1)}_{r-1}$, which are the  ($k+1$)-body
generalizations\cite{simon2006} of two-body Hamiltonians $\hat
H^{(2)}_{r-1}$ where the only non-zero two-body
pseudopotentials\cite{haldane1983} are  $V_{m} > 0$ for $m \le r-1$
However, for $k > 1$, $r > 3$ (and $(k+1)$ and $(r-1)$ relatively
prime), the number of linearly-independent  null states of $\hat
H^{(k+1)}_{r-1}$ is larger than the set of ($k$,$r$,$N$)-admissible
Jacks, and in particular, the homogeneous $\nu$ = $k/r$ FQH state
with $\lambda$ = $\lambda^0(k,r)$ is not in general unique, as seen
in Table I of \cite{simon2007}. For example, for ($k$,$r$) = (3,4),
(spinless boson states with $N$ = $3\bar N$, $N_{\Phi}$ = $4(\bar
N-1)$), the (3,4,$N$)-admissible Jack with $\lambda$ =
$\lambda^0(3,4)$ is a zero-mode eigenstate of $\hat H^{(4)}_{3}$ but
is not unique. We did not find any other local Hermitian $n$-body
pseudopotential operators that could be added to $\hat H^{(4)}_3$ to
make this Jack a unique null state, so it remains unclear how to
define the $k>1$, $r>3$ Jack FQH states as unique null states of a
model Hamiltonian (although requiring them to also be eigenstates of
the non-Hermitian operator $\mathcal H_{\rm LB}$ does make them
unique). Mathematically, they are related to
correlation functions of primary fields of non-unitary
minimal-sequence conformal field theories
\cite{feigin2002,simon2006}.

In conclusion, we have identified a number of model bosonic FQH
ground states at  $\nu$ = $k/r$ (with $k+1$ and $r-1$ relatively
prime) with a set of special Jack symmetric polynomials, whose
expansion in  monomials (free-particle occupation-number states) is
(recursively) known. The FQH states described here, being single
Jacks, are eigenstates of a multiplet of $N$ mutually commuting,
higher derivative many-body operators (Sekiguchi
operators)\cite{sekiguchi1995,bernard1993}, of which the first is
$E_1$ = $M$, the total momentum, and the second is ${\cal{H}}_{LB}
(\alpha)$. We obtained the $(k,r,N)$-admissible partitions
\cite{feigin2002} as a special case of the highest weight conditions
on the Jacks that generalizes to a $(k,r,s,N)$-admissibility to
include quasiparticle states \cite{uslater}. It may be hoped that
the identification of such larger underlying algebraic properties of
the conformally-invariant model FQH state will lead to the
development of analytic (as opposed to numerical) techniques for
calculation of their correlation functions an other properties.

This work was supported in part by the U.S. National Science Foundation (under MRSEC Grant No. DMR-0213706
at the Princeton Center for Complex Materials).

\end{document}